\documentclass[12pt]{iopart}
\usepackage{iopams} 
\input epsf
\expandafter\let\csname equation*\endcsname\relax
 \expandafter\let\csname endequation*\endcsname\relax
 \usepackage[fleqn]{amsmath}
\usepackage{amsopn}
\usepackage{epsfig}
\usepackage{subfig}
\usepackage{graphics,psfrag,rotating}
\usepackage{graphicx}
\usepackage{dcolumn}
\usepackage{bm}
\usepackage{epstopdf}
\usepackage{color}
\usepackage[usenames,dvipsnames,svgnames]{xcolor}
\usepackage[colorlinks=true,
      linkcolor=red,
      urlcolor=gray,
      citecolor=blue]{hyperref}
 \usepackage[]{cite}
\def\3nab{\tilde{\nabla}}

\def \t {\tilde}

\def\hsp5{\hspace{5mm}}

\def\case#1/#2{\textstyle\frac{#1}{#2}}

\def\ber {\begin{eqnarray}}
\def\eer {\end{eqnarray}}
\def\bea {\begin{eqnarray}}
\def\eea {\end{eqnarray}}

\def\bc {\begin{center}}
\def\ec {\end{center}}
\def\case#1/#2{\frac{#1}{#2}}

\newcommand{\bw}{\begin{widetext}}
\newcommand{\ew}{\end{widetext}}

\newcommand{\be}{\begin{equation}}
\newcommand{\bse}{\begin{subequation}}
\newcommand{\ese}{\end{subequation}}
\newcommand{\ee}{\end{equation}}
\newcommand{\eei}{\end{eqnarray}\indent\indent}
\newcommand{\ba}{\begin{array}}
\newcommand{\ea}{\end{array}}
\newcommand{\bal}{\begin{eqnarray}}
\newcommand{\eal}{\end{eqnarray}} 
\newcommand{\n}{{}^{(3)}\nabla}

\def\case#1/#2{\textstyle\frac{#1}{#2} }

\newcommand{\bver}{\begin{verbatim}}
\newcommand{\ever}{\end{verbatim}}

\begin{document}
\begin{center}
 \textbf{Cosmic hierarchy in $f(R)$ gravity}
\end{center}
\hfill\newline
Heba Sami  and Amare Abebe\\
Email: hebasami.abdulrahman@gmail.com\\
\hfill\newline
 Center for Space Research, North-West University, South Africa\\

\date{\today}

%%%%%%%%%%%%%%%%%%%%%%%%%%%%%%%%%%%%%%%%%%%%%%%%%%%%%%%%%%%%%%%%%%%%%%%%%%%%%%%%%%%%%%%%%%%%%%%%%%%%%
\begin{abstract}\\
In this contribution, we investigate hierarchical nature of large-scale structure clustering through the oscillatory nature of the solutions of the Schr\"{o}dinger-like Friedmann equation in a modified gravitational background described by the $f(R)$ gravity theory.  We find the cosmological solutions to the  Schr\"{o}dinger equation for different ranges of $n$  of the $R^{n}$ toy model for both radiation and matter-dominated epochs of the expansion history  for open, flat and closed spacetimes.  Our results show that, for certain choices of the model parameters and initial conditions, the formation and distribution of cosmic structures might indeed be hierarchical, leading to a natural explanation for the breakdown of the cosmological principle on small scales.
\end{abstract}
 
 \maketitle

%%%%%%%%%%%%%%%%%%%%%%%%%%%%%%%%%%%%%%%%%%%%%%%%%%%%%
\section{Introduction}
The structures of the Universe in its present state (galaxies, clusters and superclusters) are predicted to have formed through the constant merging of smaller structures. Observations tell us that our universe is approximately hierarchical on small scales of the order  $100h^{-1}$ Mpc and below, in which, on large scales the Universe is not homogeneous  and inconsistent with the cosmological principle.  There were many attempts to find  the simplest cosmological models for a hierarchy and to explain the idea of a hierarchical cosmology (in which the matter is indefinitely clumped) as presented in  (see \cite{de1970case, haggerty1970non, davis1992hierarchical, wesson1975relativistic, wesson1978general}, for examples).  The overall tendency of the Universe to cluster  with a decrease in redshift and the clustering properties of objects  suggest that there could be some sort of hierarchical method in which the Universe periodically evolves over time.  Within the framework of relativistic cosmology, the oscillating universe theory was introduced by Friedmann. Friedmann described the oscillation of the Universe as a single cycle from big bang to crunch,  when he analyzed  different solutions to Einstein's field equations for an isotropic and homogeneous universe for a particular case, where the cosmological constant, $\Lambda \leq 0$. He showed that the radius of curvature became a periodic function of the time \cite{friedman1922krummung}. The idea of a cyclic universe was rejected by Einstein, until, Edwin Hubble and Melvin Slipher made their analysis of the measurements of redshift and the idea of the expanding universe. Cosmology with an oscillating Hubble parameter was proposed  by Morikawa \cite{morikawa1990oscillating, morikawa1991universe}, showing that, the Hubble parameter might be responsible for the periodic distribution of galaxies and the motivation for this stem is in part, by observations from the deep narrow-cone pencil beam surveys of \cite{broadhurst1990large}  which found an excess of correlation in the galaxy distribution and suggested that the Universe may have an oscillatory nature.  Cosmological models with an oscillatory behaviour have been studied by many groups,  in \cite{durrer1996oscillating}; it has been shown that the oscillating universe can provide some solutions to the flatness, horizon and the entropy problems of the standard cosmology. Extending the work done in \cite{namane2018oscillating, NeoNamane} and following Capozziello \cite{capozziello2000oscillating} and Rosen \cite{rosen1993quantum},  this particular piece of literature is aimed at confirming the oscillatory clustering of a universe that is described from the radiation era to today's present epoch in terms of one of the modified theories of gravity,  $f(R)$ gravity.  This inclusive periodic structure of the variational recapitulation of the cosmos is representative of the wave-like nature of the Universe that is ideally mimicked by the Schr\"{o}dinger equation on non-relativistic scales.  An investigation of the power model is made against a quantum approximation that is evaluated alongside different equation of state parameters (radiation and matter). Oscillations in redshift can be considered as some sort of quantization \cite{capozziello2000oscillating, tifft1976discrete} and all quantities containing $H$ or $z$ have to oscillate, given the cosmological scale factor $a(t)$ and redshift $z$, we have the Hubble parameter $H$ given by
\begin{eqnarray}\label{zz}
\frac{\dot{a}}{a}=\frac{\Theta}{3}\equiv H=-\frac{\dot{z}}{1+z}\;.
\end{eqnarray}
\newline
This paper is organised as follows: in Sections \ref{Sec1},  we review the field equations in $f(R)$ theory of gravity and we get  the Friedmann equation that describes a modified gravitational background in terms of $f(R)$ gravity. In Sections \ref{Sec2} and \ref{Sec3}, respectively,  we construct the  cosmological Schr\"{o}dinger equation and we get a solutions for such equation by considering $R^n$ models. Section \ref{sec3} is devoted for discussions and conclusions.
\section{Field equations}\label{Sec1}
The cosmological evolution equations of GR come from the Einstein field equations
\begin{eqnarray}
R_{\mu \nu} -\dfrac{1}{2} g_{\mu \nu} R = \kappa T_{\mu \nu}\;,
\end{eqnarray}
where $R_{\mu \nu}$ is the Ricci tensor, $R$ is the Ricci scalar, $g_{\mu \nu}$ is the metric tensor, $\kappa= \dfrac{8\pi G}{c^{4}}$ is a constant known as the Einstein gravitational constant, $G$ is Newton’s gravitational constant, $c$ is the speed of light in vacuum and  $T_{\mu \nu}$ is the stress energy tensor. Such equations can be derived from the Hilbert-Einstein action
\begin{eqnarray}
S_{EH}= \dfrac{1}{2\kappa}\int  \sqrt{-g}\Big(R+2\mathcal{L}_{m}(g_{\mu \nu},\psi)\Big)d^{4}x\;.
\end{eqnarray}
Here $\mathcal{L}_{m}$ is the Lagrangian density relative to the matter fields.
In the case of $f(R)$ gravity theory, the action is given by \cite{joras2011some, sotiriou2006f, sotiriou2010f, sotiriou2007modified,clifton2012modified,vitagliano2010dynamics,capozziello2011extended}
\begin{eqnarray}\label{FRA}
S_{f(R)}= \dfrac{1}{2\kappa}\int \sqrt{-g}\Big(f(R)+2\mathcal{L}_{m}(g_{\mu \nu},\psi)\Big)  d^{4}x\; .
\end{eqnarray}
and the corresponding generalized field equations are given by
\begin{eqnarray}\label{metricFE}
 f'(R)R_{\rho \nu} -\dfrac{1}{2}f(R)g_{\rho \nu}-\Big(\nabla_{\rho} \nabla_{\nu}-g_{\rho \nu} \nabla_{\rho}\nabla^{\rho}\Big)f'(R)= \kappa T_{\rho \nu}\; ,
\end{eqnarray}
where  $f'(R)= \dfrac{df}{dR}$ and $\nabla_{\rho}$ is the standard covariant derivative which is formed  via the usual Levi-Civita connection. The energy-momentum tensor of matter is given as
\begin{eqnarray}\label{emt}
T_{\rho \nu}= -\dfrac{2\delta\left(\sqrt{-g}\mathcal{L}_{m}\right)}{\sqrt{-g} \delta g^{\rho \nu}}\; .
\end{eqnarray}
From the last two terms of Eqs. \eqref{metricFE}, we notice that the field equations obtained in $f(R)$ are of fourth-order partial differential equations in the metric $g_{\rho \nu}$. However, the fourth-order terms vanish when $f'$ is a constant, i.e. for an action which is linear in  $R$. Thus, it is straightforward for these equations to reduce to the Einstein equations once $f(R)= R$ (reduces to GR) \cite{sotiriou2007modified}. The trace of Eq. \eqref{metricFE} is given by \cite{sotiriou2010f}
\begin{eqnarray}\label{trace}
f'(R) R -2f(R) + 3 \nabla^{\rho}\nabla_{\rho} f'(R) = \kappa T\;.
\end{eqnarray}
 As a result of the extra degrees of freedom provided by the curvature fluid induced by the modified Ricci scalar,  the total  effective energy density and an isotropic pressure of the standard matter and curvature fluid are defined as
\begin{eqnarray}\label{1}
&&\mu = \dfrac{\mu_{m}}{f'} + \mu_{R}\;,\quad\quad p = \dfrac{p_{m}}{f'} + p_{R}\;, \\
&&\label{3}  p_{R} \equiv  \dfrac{1}{f'}\Bigg(\dfrac{1}{2}\big(f - Rf'\big) + f^{\prime\prime}\ddot R + f''\dot R^{2} +\dfrac{2}{3} \theta f''\dot R\Bigg)\;,\\
&& \label{4}  \mu_{R}\equiv  \dfrac{1}{f'}\Bigg(\dfrac{1}{2}\big(Rf' - f\big) - \theta f''\dot R\Bigg)\;.
\end{eqnarray}
The fluid conservation equations are given by
\begin{eqnarray}
&&  \dot \mu_{m} + 3H(\mu_{m} + p_{m})=0\;,\\
&& \dot \mu_{R} + 3H(\mu_{R} + p_{R}) - \mu_{m}\dfrac{f'' \dot R}{f'^2} = 0\;,
\end{eqnarray}
whereas the generalized  Raychaudhuri equation is given as
\begin{eqnarray}\label{Fri}
&&\dfrac{\ddot{a}}{a}= -\dfrac{4\pi G}{3}(\mu +\dfrac{3p}{c^{2}})\;.
\end{eqnarray}
By substituting Eqs. \eqref{1}  into Eq. \eqref{Fri}, we get
\begin{eqnarray}\label{Fri1}
&& \dfrac{\ddot{a}}{a}= -\dfrac{4\pi G}{3}\Big(\dfrac{\mu_{m}}{f'}+ \mu_{R}+ \dfrac{3p_{m}}{f'c^{2}}+\dfrac{3p_{R}}{c^{2}}\Big)\;.
\end{eqnarray}
From Eqs. \eqref{3} and \eqref{4}, it is possible to write the preceding equation as
\begin{eqnarray}\label{Fri2}
&&\dfrac{\ddot{a}}{a}= -\dfrac{4\pi G}{3} \Big(\dfrac{\mu_{m}}{f'}+\dfrac{1}{2f'}(Rf' -f)-\dfrac{\theta f''\dot{R}}{f'}+\dfrac{3p_{m}}{f'c^{2}}- \dfrac{3}{2f' c^{2}}(Rf'-f) \nonumber\\
 &&\quad\quad+ \dfrac{3f''\ddot{R}}{f'c^{2}}+\dfrac{3f'''\dot{R}^{2}}{f'c^{2}}+\dfrac{2\theta f''\dot{R}}{f'c^{2}}\Big)\;.
\end{eqnarray}
Further simplification of Eq. \eqref{Fri2} using the trace equation
\begin{eqnarray}\label{trace1}
R= \frac{8\pi G}{c^{2}}(\mu -\frac{3p}{c^{2}})\;.
\end{eqnarray}
leads to 
\begin{eqnarray}\label{eq:14}
\dfrac{\ddot{a}}{a}= -\dfrac{4\pi G}{3f^{'}}\Big( 2\mu_{m}+(Rf'- f) -2\theta f^{''}\dot{R}\Big) + \frac{Rc^{2}}{6}\;.
\end{eqnarray}
It is also possible to rewrite the Friedmann equations as
\begin{eqnarray}\label{Ray1}
&&H^{2}= \dfrac{4\pi G}{3f'}\Big(2\mu_{m} +(Rf' -f)-2\theta f''\dot{R}\Big)- \dfrac{k c^{2}}{a^{2}}\;,\\
&&\dot{H}+ H^{2}=-\dfrac{4\pi G}{3f'}\Big( \mu+ \frac{3p}{c^{2}}\Big)\;.
\end{eqnarray}
In addition, it is observed that from Eq. \eqref{eq:14}, Eq. \eqref{Ray1} can be written as 
\begin{eqnarray}\label{Ray11}
H^{2}= \dfrac{Rc^{2}}{6}- \dfrac{\ddot{a}}{a}-\dfrac{k c^{2}}{a^{2}} \;. 
\end{eqnarray}
The probability of attaining the actual dynamics of an idealistic universe that mimics particles governed by quantum fields is subject to the collapse of the wave-function that describes it. Given this fact, it is imperative for one to first determine the potential that is responsible for such evolutionary manifestation. The potential and kinetic energy are thus attained in the following way by the  introduction of a mass term which is said to describe a Universe \cite{rosen1993quantum} or the particle-like dynamics of galaxies \cite{capozziello2000oscillating} within the context of the conservation of mechanical energy.
\section{The cosmological Schr\"{o}dinger equation }\label{Sec2}
From Eq. \eqref{Ray11}, it can be deduced that the kinematics of the system can be quantified upon the inference of the existence of a mass that propels its dynamical evolution. As such, an alternate representation of the equation is given as
\begin{eqnarray}\label{Ray2}
 \dfrac{1}{2}m\dot{a}^{2}-\dfrac{1}{2}\Big( \dfrac{Rc^{2}}{6} +\dfrac{\ddot{a}}{a}\Big)ma^{2}=-\dfrac{1}{2}mk c^{2}\;.
\end{eqnarray}
Here, the cosmological constant that is present in the general theory of relativity in order to give an account for the existence of dark energy is replaced by an expression that is in terms of the geometry of the modified gravitational manifold. In \cite{capozziello2000oscillating}, the potential acquired was that of a point-like galaxy found within the Universe. As a result, the same approximation is made here so as to have a similar account of the subsequent potential that drives the resulting wave-function. Motivation for this stems from the perceived periodic nature of all physically observed occurrences that seem to rely on what is known as a galaxy-galaxy correlation function that is presented as
\begin{eqnarray}
 \xi_{j}=\dfrac{C_{j}}{r^{\rho_{j}}}\;.
\end{eqnarray}
According to work done in \cite{busarello1994apparently}, projected distances that are observed in the Universe may be modelled through the use of this power law. The subscript $j$ is representative of the different cosmological objects that are as a result of clustering while $C_{j}$ and $\rho_{j}$ articulate the correlation amplitudes and indexes of the slope respectively.

Let us now identify the potential energy and total mechanical energy terms as
\begin{eqnarray}\label{vv}
 &&V(a) = -\dfrac{1}{2} \Big( \dfrac{Rc^{2}}{6}+\dfrac{\ddot{a}}{a}\Big)ma^{2}\;,\\
&&E = -\dfrac{1}{2}mkc^{2}\;,
\end{eqnarray}
whereas the first term in Eq. \eqref{Ray2} is taken as the kinetic energy term.
Thus, the Schr\"{o}dinger equation is a suitable relation that is adequate enough for the purpose of determining evolutionary states of the wave-function and the inherent eigensolutions that are duly presented. However, constructing a dynamical system whose spatial dependence lies in the progression of the scale factor necessitates that the Schr\"{o}dinger equation be given as
\begin{eqnarray}\label{shcro0}
 i \hbar \dfrac{\partial \Psi}{\partial t} = -\dfrac{\hbar^{2}}{2m}\dfrac{\partial^{2} \Psi}{\partial a^{2}} + V(a)\Psi\;,
\end{eqnarray}
where $\Psi= \Psi(a,t)$. By following ordinary quantum mechanics, a stationary state of energy $E$ is 
\begin{eqnarray}
\Psi(a,t)= \psi(a)e^{-iEt/\hbar}\;, 
\end{eqnarray}
and the Schr\"{o}dinger stationary equation is
\begin{eqnarray}
\Big[ E-V(a)\Big]\psi= -\dfrac{\hbar^{2}}{2m}\dfrac{\partial^{2}\psi(a)}{\partial a^{2}}\;.
\end{eqnarray}
The second-order differential equation that presents itself is a step closer towards getting a relation for the wave-function that describes the point-like mass that oscillates against an $f(R)$ gravitational background, i.e
\begin{eqnarray}\label{psi}
 \psi''= -\dfrac{2m}{\hbar^{2}}\Big[E-V(a)\Big]\psi\;.
\end{eqnarray}
In looking at Eq. (\ref{Ray1}) for an expression of the scale factor's temporal derivative, the following is attained
\begin{eqnarray}\label{psi0}
\psi''=  -\dfrac{m^{2}}{\hbar^{2}}\Big[-k c^{2}+\dfrac{4\pi G }{3f'}\Big( 2\rho_{m}+(Rf'-f)-2\theta f''\dot{R}\Big)a^{2}\Big]\psi\;.
\end{eqnarray}
If we let $A= m^{2}$, $4\pi G= \dfrac{1}{2}$, $E_T =-k$, $\hbar= c=1$ and \\
$U(a)=- \dfrac{4\pi G }{3f'}\Big( 2\rho_{m}+(Rf'-f)-2\theta f''\dot{R}\Big)a^{2} \;,$\\
Eq. (\ref{psi0}) reduces to 
\begin{eqnarray}\label{psi03}
\psi''+ A\Big(E_T - U(a)\Big) \psi=0\;,
\end{eqnarray}
where $U(a)$ is the potential energy,  $E_T$ is the total energy, $m$  is the mass of a galaxy and $|\psi|^2 $ is the probability to find such a galaxy at a given $a(t)$.
\begin{itemize}
\item [1.] If we are in some regime where $E_T >> U(a)$, with $E_T$ being positively defined ($k=-1$), Eq. (\ref{psi03}) has a solution of the form
\begin{eqnarray}
\psi(a)= C_{1} \sin (\sqrt{ A E_T}a) + C_{2} \cos(\sqrt{A E_T}a)\;.
\end{eqnarray}
\item [2.] If $E_T << U(a)$, with $E_T$ being negatively defined ($k=1$), Eq. (\ref{psi03}) has a solution of the form
\begin{eqnarray}
\psi(a)= C_{1} e^{\sqrt{A E_T} a} + C_{2} e^{\sqrt{A E_T}a}\;.
\end{eqnarray}
\end{itemize}
From these solutions it is easy to see that oscillatory behaviours are recovered. We also pass from $a(t)$ to the observable redshift $z$, from Eq. \ref{zz} and as a result
\begin{eqnarray}
\psi' = \dfrac{\partial\psi}{\partial a}= \dfrac{\partial \psi}{\partial z}\dfrac{\partial z}{\partial a} = -\dfrac{(1+z)^{2}}{a^{2}_{0}}\dfrac{\partial \psi}{\partial z}\;, 
\end{eqnarray}
\begin{eqnarray}
\psi'' = \dfrac{\partial}{\partial a}\Big(-\dfrac{(1+z)^{2}}{a^{2}_{0}}\dfrac{\partial \psi}{\partial z} \Big)= -\dfrac{(1+z)^{2}}{a^{2}_{0}} \dfrac{\partial^{2} \psi}{\partial z^{2}} \dfrac{\partial z}{\partial a} - \dfrac{\partial \psi}{\partial z}\big( \dfrac{2(1+z)}{a_{0}}\big) \dfrac{\partial z}{\partial a}\;,
\end{eqnarray}
and we get
\begin{eqnarray}
\psi'' = \dfrac{(1+z)^{4}}{a^{2}_{0}}\dfrac{\partial^{2} \psi}{\partial z^{2}}+ \dfrac{2(1+z)^{3}}{a^{2}_{0}} \dfrac{\partial \psi }{\partial z}\;.
\end{eqnarray}
Therefore Eq. \eqref{psi0} is written as
\begin{eqnarray}\label{psi3}
\dfrac{\partial^{2} \psi}{\partial z^{2}}+ \dfrac{2}{(1+z)}\dfrac{\partial \psi }{\partial z}=-\dfrac{m^2}{\hbar^{2} (1+z)^{4}}  \Big(-c^{2}k+\dfrac{4\pi G a^{4}}{3f'(1+z)^{2}}\Big( 2\rho_{m}+(Rf'-f)-2\theta f''\dot{R}\Big)\Big)\psi\;.\nonumber\\
\end{eqnarray}
Eq. (\ref{psi3}) reduces to 
\begin{eqnarray}\label{psi0003}
\psi''+\dfrac{2}{(1+z)}\psi' + \frac{A}{ (1+z)^{4}}\Big(E_T - U(z)\Big) \psi=0\;,
\end{eqnarray}
where
$U(z)=- \dfrac{4\pi G}{3f'(1+z)^{2}}\Big( 2\rho_{m}+(Rf'-f)-2\theta f''\dot{R}\Big) \;.$\\
In the following section, we will find the solutions of this wave equation in both matter- and radiation-dominated eras.

\section{Solutions}\label{Sec3}
In order to study the oscillatory natures of the Schr\"{o}dinger equation  \eqref{psi0003}, we consider here the  $R^n$ model, one of the $f(R)$ toy models considered to be the simplest and widely studied form of $f(R)$ gravitational theories. The Lagrangian density of such models is given as
\begin{eqnarray}
f(R)= \beta R^{n}\;,
\end{eqnarray} 
 where $\beta$ represents the coupling parameter and the exponent $n\neq1$ for non-GR cosmological models.  Furthermore, we consider the scale factor solution of the form \cite{carloni2005cosmological}
\begin{eqnarray}
a= a_{0}\Big(\dfrac{t}{t_0}\Big)^{\dfrac{2n}{3(1+w)}}\;,
\end{eqnarray} 
and one can obtain the following expressions for the expansion, the Ricci scalar and the effective matter energy density respectively \cite{carloni2005cosmological}:
\begin{eqnarray}\label{10}
&&\theta= \dfrac{2n}{(1+w)t}\;,
  \\&& \label{110} R = \frac{4n(4n-3(1+w))}{3(1+w)^{2}t^{2}}\;, \\&& \label{1100}
\mu_{m} = n\Big(\dfrac{3}{4}\Big)^{(1-n)} \Big( \dfrac{n(4n-3(1+w))}{(1+w)^{2}t^{2}}\Big)\times \nonumber\\
&&\Big( \dfrac{4n^{2}-2(n-1)(2n(3w+5)-3(1+w))}{3(1+w)^{2}t^{2}}\Big)\;.
\end{eqnarray}
In the radiation-dominated era, these solutions correspond to 
\begin{eqnarray}\label{expansion1}
&&\Theta= \dfrac{3n}{2t}\;, \quad R=\dfrac{3n(n-1)}{t^{2}}\; ,\\
&&\label{matterenergy1}
\mu_{r}=3n \left(\dfrac{3n(n-1)}{t^{2}}\right)^{(n-1)} \Big( \dfrac{(-5n^{2}+8n -2)}{4t^{2}}\Big)\;,
\end{eqnarray}
whereas for the matter-dominated era of the Universe's evolution,  Eqs. (\ref{10}-\ref{1100}) reduce to:
\begin{eqnarray}\label{expansion2}
&&\Theta= \dfrac{2n}{t}\;, \quad R=\dfrac{4n(4n-3)}{3t^{2}}\; ,\\
&&\label{matterenergy2}
\mu_{d}=n \left(\dfrac{4n(4n-3)}{3t^{2}}\right)^{(n-1)} \Big( \dfrac{(-16n^{2}+26n -6)}{3t^{2}}\Big)\;.
\end{eqnarray}
\subsection{Solutions for the radiation-dominated era}
 In this subsection, we get the solution for the  the Schr\"{o}dinger  equation Eq. \eqref{psi0003} in the radiation-dominated era. By using the expressions for the expansion, the Ricci scalar and the effective matter energy density Eqs. \eqref{expansion1}-\eqref{matterenergy1}, we get
\begin{eqnarray}
&&\dfrac{\theta f'' \dot{R}}{f'}= \dfrac{-12 H^{2}(n-1)}{n}\;,\\&&
\dfrac{1}{f'}(Rf' -f)= \dfrac{12H^{2}(n-1)^2}{n^{2}}\;,\\&&
\dfrac{\rho_{r}}{f'}= \dfrac{3 H^{2}(-5n^{2}+8n-2)}{ n^{2}}\;,
\end{eqnarray}
where 
$H^{2}= \frac{4n^2}{9}(1+z)^{3/n} \;.$ Therefore, we have Eq. (\ref{psi0003}) with \\
$U(z)=- \dfrac{-n^{2} }{4(1+z)^{2-4/n}} \Big(\dfrac{(-5n^2+8n-2)}{n^2}+\dfrac{2(n-1)^2}{n^2} +\dfrac{4(n-1)}{n}\Big) \;.$ 
\subsection*{\bf{For  the case when $k=0$}}
\begin{figure}[h!]
\begin{minipage}[b]{0.5\linewidth}
\includegraphics[width=0.9\textwidth]{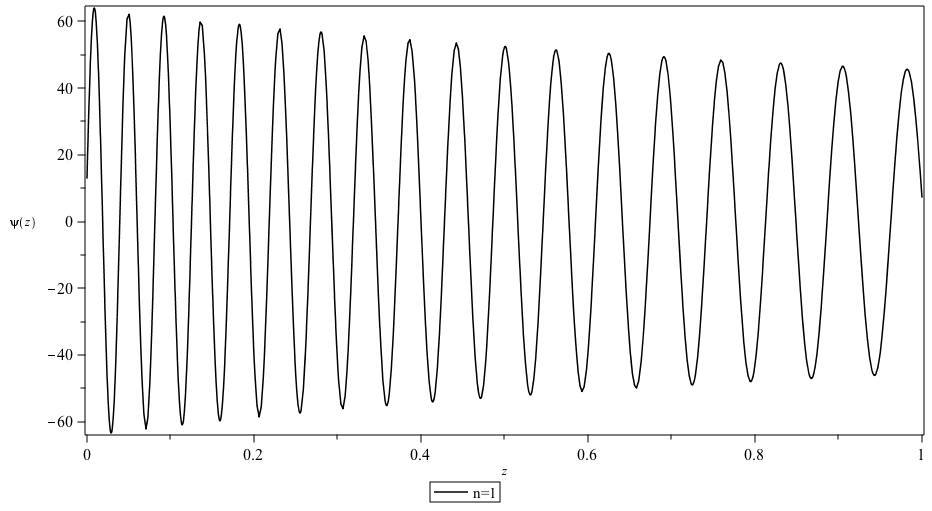}
\caption{$\psi(z)$ versus $z$ in the radiation-dominated epoch  for $A= 10^5$, $k=0$ and $n$= 1.}
\label{Figure 1}
\end{minipage}
\qquad
\begin{minipage}[b]{0.5\linewidth}
\includegraphics[width=0.9\textwidth]{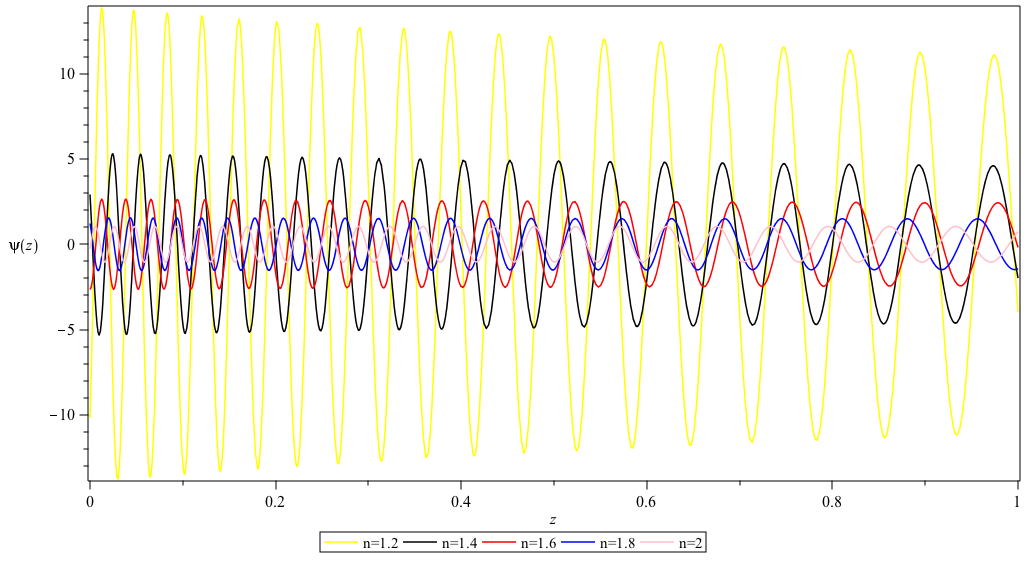}
\caption{$\psi(z)$ versus $z$ in the radiation-dominated epoch  for $A= 10^5$, $k=0$ and $1.1 \leq n \leq 2$.}
\label{Figure 2}
\end{minipage}
\qquad
\begin{minipage}[b]{0.5\linewidth}
\includegraphics[width=0.9\textwidth]{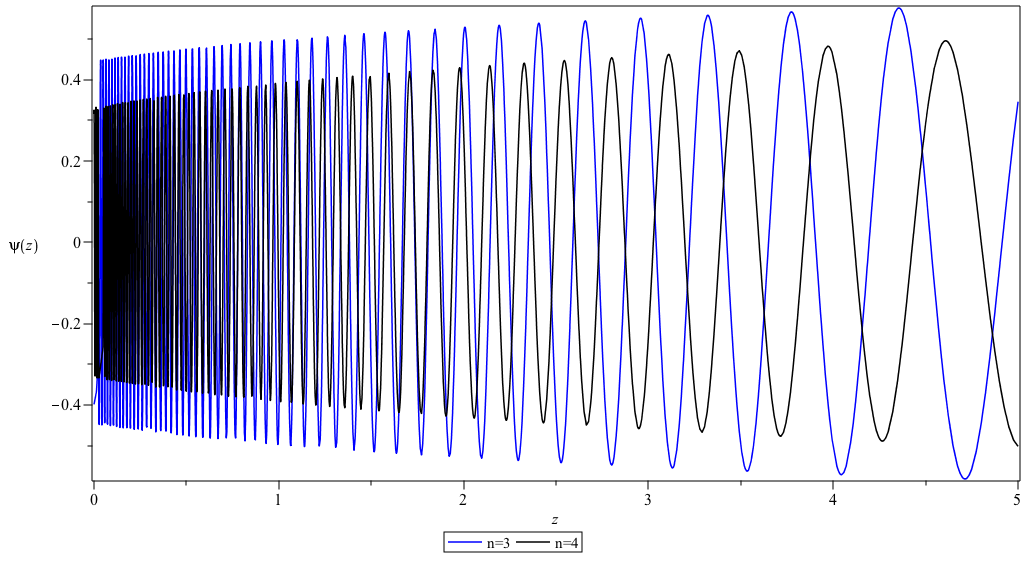}
\caption{$\psi(z)$ versus $z$ in the radiation-dominated epoch  for $A= 10^5$, $k=0$ and $n \geq 3$.} 
\label{Figure 3}
\end{minipage}
\qquad
\begin{minipage}[b]{0.5\linewidth}
\includegraphics[width=0.91\textwidth]{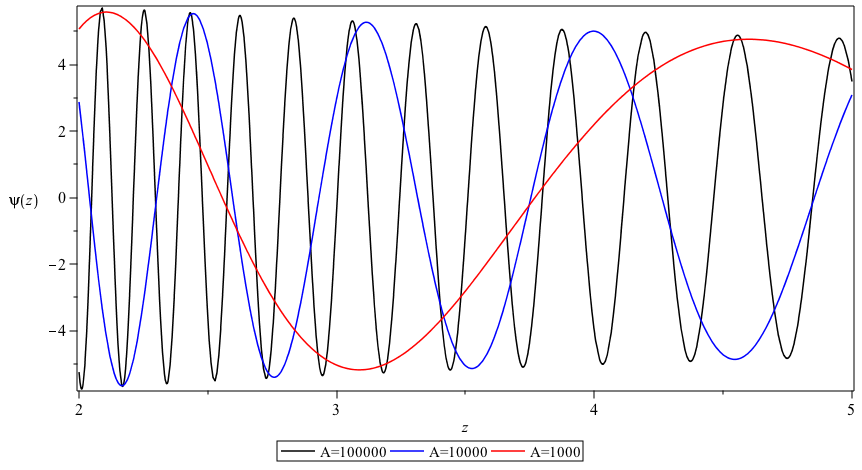}
\caption{$\psi(z)$ versus $z$ in the radiation-dominated epoch  for $n= 1.3$, $k=0$.} 
\label{Figure 4}
\end{minipage}
\end{figure}

For  the case when $k=0$,   Eq. (\ref{psi0003}) is written as
\begin{eqnarray}
\psi''+\dfrac{2}{(1+z)}\psi' - \frac{A}{ (1+z)^{4}}  U(z) \psi=0\;,
\end{eqnarray} 
the general solution is combination for Hypergeometric functions 
\begin{eqnarray}
&&\psi(z)= C_{1} \mathcal{H} \Big([], [\dfrac{(3n-4)}{(4n-4)}], \dfrac{An^{2} D (1+z)^{\frac{-4n+4}{n}}}{16(n-1)^2}\Big)+ \nonumber\\&&C_{2} \dfrac{1}{(1+z)}\mathcal{H} \Big([], [\dfrac{(5n-4)}{(4n-4)}], \dfrac{An^{2}D (1+z)^{\frac{-4n+4}{n}}}{16(n-1)^2}\Big)\;,
\end{eqnarray}
where $$D=- \dfrac{-n^{2} }{4} \Big(\dfrac{(-5n^2+8n-2)}{n^2}+\dfrac{2(n-1)^2}{n^2} +\dfrac{4(n-1)}{n}\Big)\;. $$\\
 %and $\mathcal{J}_{\nu}(a)$ and $\mathcal{Y}_{\nu}(a)$ 
 %in the solution to Besselâs equation are referred to as  Bessel functions of the first and the second kind respectively.
  For $n=1$, the general solution is given as
\begin{eqnarray}
\psi(z)= C_{1} z^{\Big(\dfrac{-1}{2}+ \dfrac{\sqrt{1-A}}{2}\Big)}+ C_{2} z^{\Big(\dfrac{-1}{2}- \dfrac{\sqrt{1-A}}{2}\Big)}\;.
\end{eqnarray}
Here $C_{1}$ and $C_{2}$ can be found by setting initial conditions, $\psi(z_{in})= 10^{-3}$ and $\psi'(z_{in})=0$, where $\psi(z_{in})$ is the initial value of $\psi(z)$ at $z_{in}$,  where the subscript in refers to the initial scale factor $z_{in}=2000$.  We present the results for different values of $n$ in Figs. \ref{Figure 1} - \ref{Figure 4}  for a specific choice of the (normalized) mass parameter $A= 10^5$. In Fig. \ref{Figure 1},  the GR result is recovered.   An oscillatory behaviour is observed and  the probability of finding a galaxy of mass $A=10^5$  increases with decreasing the redshift $z$  for values of $1\leq n \leq 2$   as in Fig. \ref{Figure 2}. While  In Fig. \ref{Figure 3}, the probabilities increase with increasing  the redshift   for values of $n \geq 3$. For a fixed  value of $n$ and different  values of the mass $A$,  see Fig.  \ref{Figure 4}, the rate of oscillations and the probabilities of finding the galaxies of different masses  are increasing  with  the mass $A$. 
\newline
\newline
\subsection*{\bf{ For  the case when $k=1$ }}
For the case when $k=1$, Eq. (\ref{psi0003}) is written as
\begin{equation}
\psi''+\dfrac{2}{(1+z)}\psi' + \frac{A}{ (1+z)^{4}}(-1-  U(z)) \psi=0\;,
\end{equation} 
and  for $n=1$, the general solution is a combination of Whittaker functions
\begin{equation}
\psi(z)=C_{3}  \mathcal{Y}_{1} \Big(0, \frac{  \sqrt{1-A}}{2}, \frac{2\sqrt{A}}{(1+z)}\Big)+ C_{4} \mathcal{Y}_{2}  \Big(0, \frac{  \sqrt{1-A}}{2}, \frac{2\sqrt{A}}{(1+z)}\Big)\;.
\end{equation}
We present the numerical solutions for different values of $n$ in the following Figs. \ref{Figure 5} - \ref{Figure 8}. We noticed the oscillatory behaviour  for $n=1$  only when the range of redshift begins at $ z \geq 0.69$ as shown in Fig. \ref{Figure 5}. While for values of $1< n \leq2 $, oscillations  observed only when the range of redshift begins at $ z \geq 1$ as shown in Figs. \ref{Figure 6} and \ref{Figure 8}. From Fig. \ref{Figure 6}, the rate of the oscillations and the amplitudes or the probabilities of finding a galaxy of mass $A= 10^{5}$,  are all decreasing with increasing  the values of $n$. For $n\geq 3$, oscillatory behaviour is observed only for small range of the redshift as presented in Fig.  \ref{Figure 7}. We also presented the numerical results for different masses $A$ in Fig. \ref{Figure 8}.
\begin{figure}[h!]
\begin{minipage}[b]{0.45\linewidth}
\includegraphics[width=0.9\textwidth]{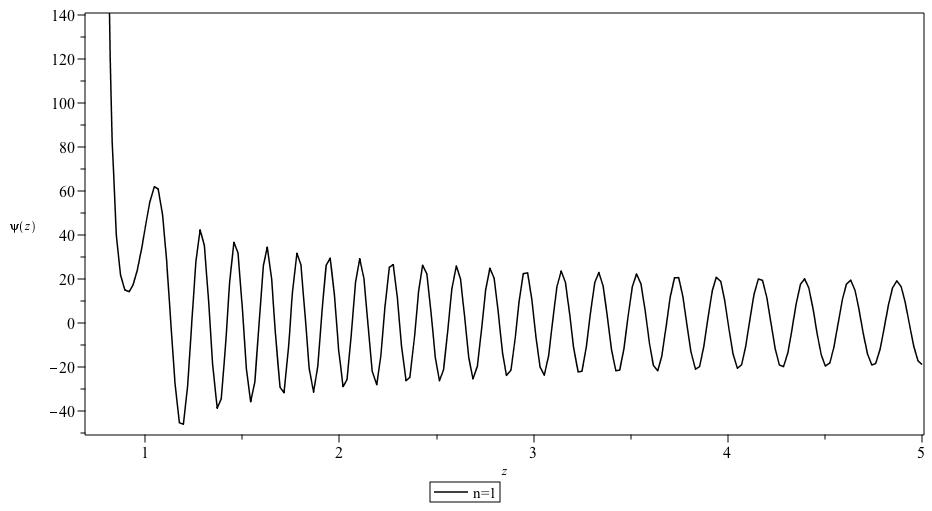}
\caption{$\psi(z)$ versus $z$ in the radiation-dominated epoch for $A= 10^5$, $k=1$ and $n=1$.}
\label{Figure 5}
\end{minipage}
\qquad
\begin{minipage}[b]{0.45\linewidth}
\includegraphics[width=0.9\textwidth]{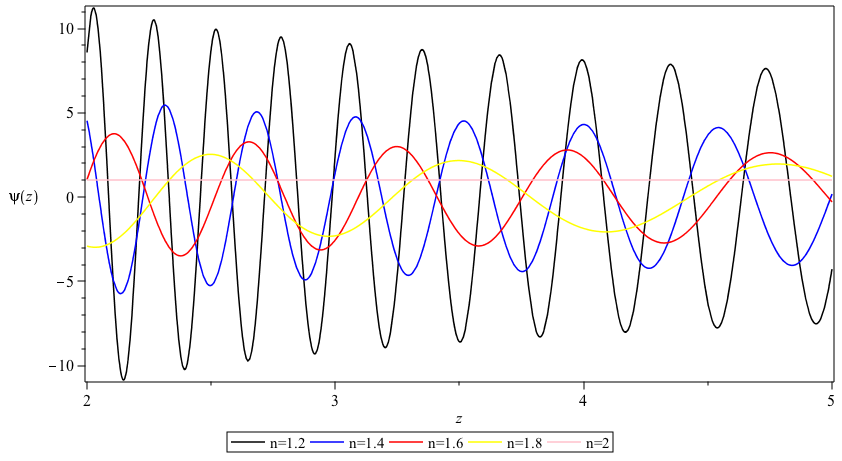}
\caption{$\psi(z)$ versus $z$ in the radiation-dominated epoch  for $A= 10^5$,  $k=1$ and $ 1< n\leq 2$.}
\label{Figure 6}
\end{minipage}
\end{figure}
\begin{figure}[h!]
\begin{minipage}[b]{0.45\linewidth}
\includegraphics[width=0.9\textwidth]{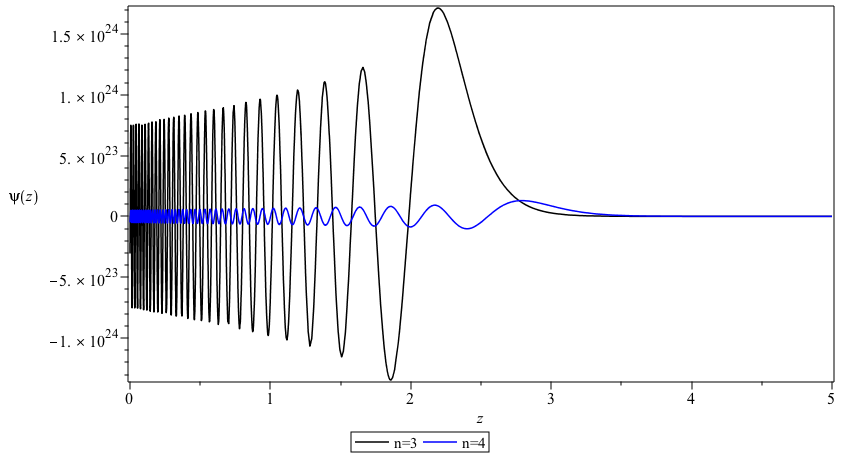}
\caption{$\psi(z)$ versus $z$ in the radiation-dominated epoch  for $A= 10^5$,  $k=1$ and $n\geq 3$.}
\label{Figure 7}
\end{minipage}
\qquad
\begin{minipage}[b]{0.45\linewidth}
\includegraphics[width=0.9\textwidth]{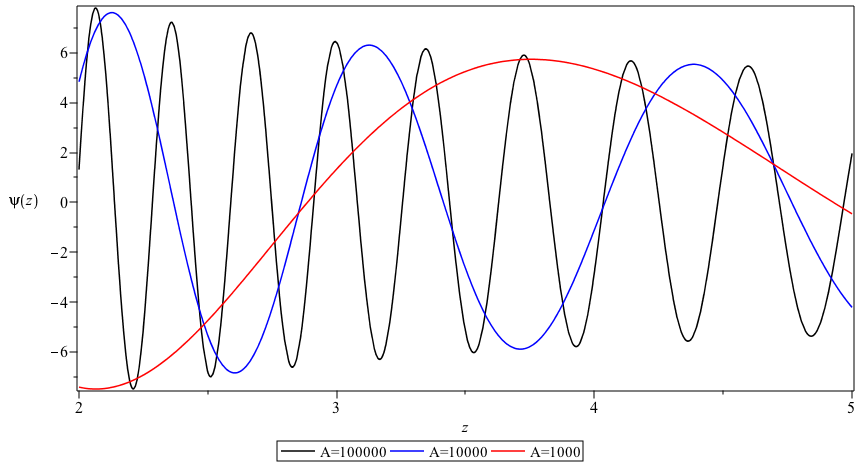}
\caption{$\psi(z)$ versus $z$ in the radiation-dominated epoch  for  $k=1$ and $n=1.3$.}
\label{Figure 8}
\end{minipage}
\end{figure}
\newpage
\subsection*{\bf{ For  the case when $k=-1$ }}
For the case when $k=-1$, Eq. \eqref{psi0003} is written as
\begin{equation}
\psi''+\dfrac{2}{(1+z)}\psi' + \frac{A}{ (1+z)^{4}}(1-  U(z)) \psi=0\;,
\end{equation} 
and  for $n=1$, the general solution is a combination of Whittaker functions
\begin{equation}
\psi(z)=C_{5}  \mathcal{Y}_{1} \Big(0, \frac{  \sqrt{1-A}}{2}, \frac{2I\sqrt{A}}{(1+z)}\Big)+ C_{6} \mathcal{Y}_{2}  \Big(0, \frac{  \sqrt{1-A}}{2}, \frac{2I\sqrt{A}}{(1+z)}\Big)\;.
\end{equation}
We present the numerical solutions for $k=-1$ and different values of $n$  in the following Figs. \ref{Figure 9} - \ref{Figure 10}. 
\begin{figure}[h!]
\begin{minipage}[b]{0.48\linewidth}
\includegraphics[width=0.9\textwidth]{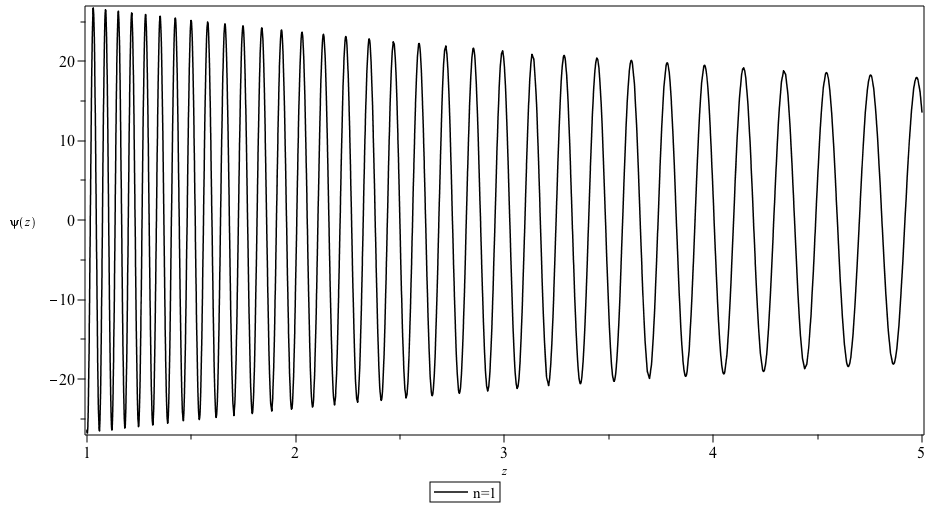}
\caption{$\psi(z)$ versus $z$  in the radiation-dominated epoch for $A= 10^5$,  $k=-1$ and $n=1$.} 
\label{Figure 9}
\end{minipage}
\qquad
\begin{minipage}[b]{0.45\linewidth}
\includegraphics[width=0.9\textwidth]{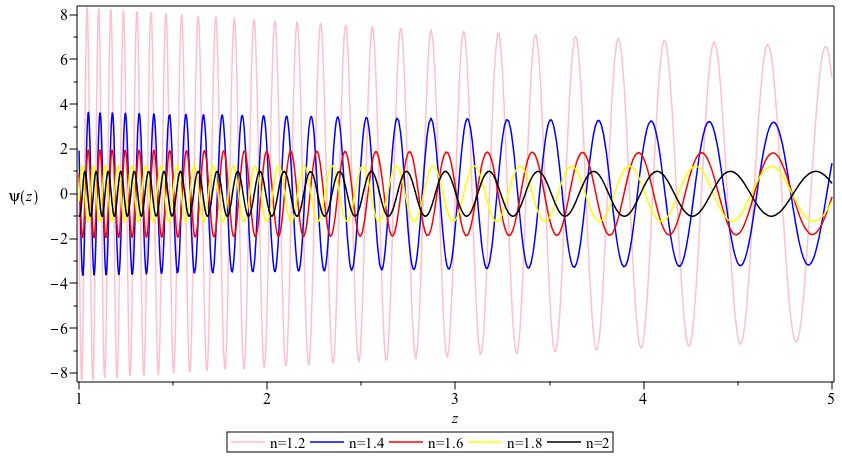}
\caption{$\psi(z)$ versus $z$ in the radiation-dominated epoch  for $A= 10^5$, $k=-1$ and $1.1 \leq n \leq 2 $.} 
\label{Figure 10}
\end{minipage}
\end{figure}
\newpage
\subsection{Solutions for the matter-dominated era}
 In this subsection, we get the solution for the  the Schr\"{o}dinger  equation Eq. \eqref{psi0003} in the matter-dominated era. From the expressions for the expansion, the Ricci scalar and the effective matter energy density Eqs. \eqref{expansion2}-\eqref{matterenergy2}, we get
\begin{eqnarray}
&&\dfrac{\theta f'' \dot{R}}{f'}= \dfrac{-9 H^{2}(n-1)}{n}\;,\\ &&
\dfrac{1}{f'}(Rf' -f)= \dfrac{3H^{2}(n-1)(4n-3)}{n^{2}}\;,\\&&
\dfrac{\rho_{d}}{f'}= \dfrac{3 H^{2}(-16n^{2}+26n-6)}{ 4n^{2}}\;.
\end{eqnarray}
Therefore, the Schr\"{o}dinger equation Eq. \eqref{psi0003}
$$U(z)= \dfrac{-2n^{2} }{9(1+z)^{2-3/n}} \Big(\dfrac{(-16n^2+26n-6)}{2n^2}+\dfrac{(n-1)(4n-3)}{n^2} +\dfrac{6(n-1)}{n}\Big)  \;.$$\\
\subsection*{\bf{ For  the case when $k=0$}}
\begin{figure}[h!]
\begin{minipage}[b]{0.5\linewidth}
\includegraphics[width=0.9\textwidth]{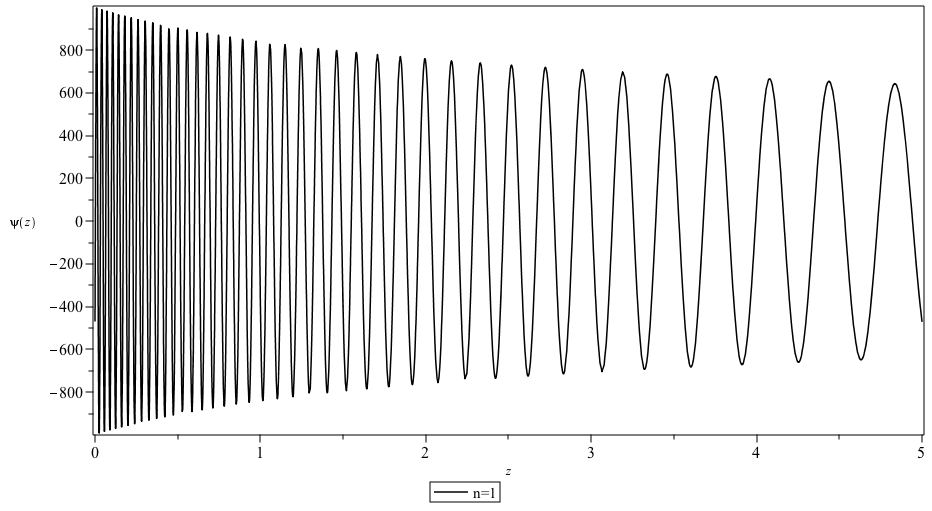}
\caption{$\psi(z)$ versus $z$ in the matter-dominated era for $A= 10^5$ and $k=0$.}
\label{Figure 14O}
\end{minipage}
\qquad
\begin{minipage}[b]{0.5\linewidth}
\includegraphics[width=0.9\textwidth]{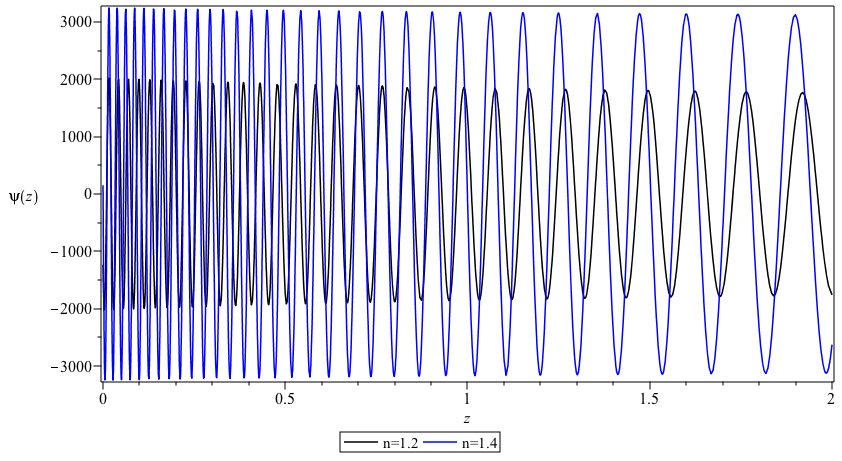}
\caption{$\psi(z)$ versus $z$   in the matter-dominated era for $A= 10^5$  $n>1 $ and $k=0$.}
\label{Figure 16O}
\end{minipage}
\end{figure}

\begin{figure}[h!]
\begin{minipage}[b]{0.55\linewidth}
\includegraphics[width=0.9\textwidth]{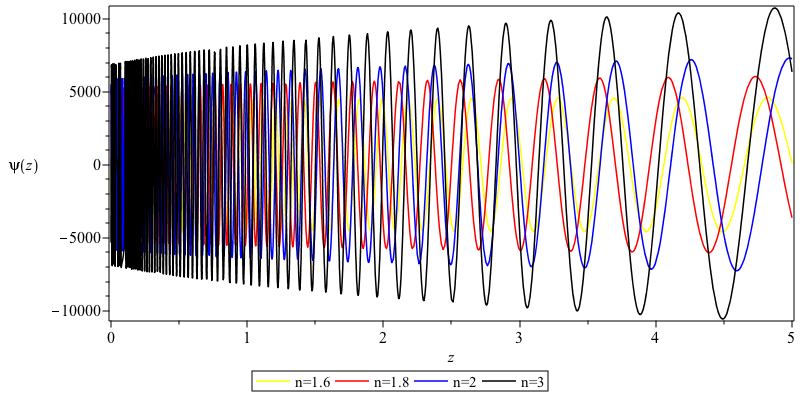}
\caption{$\psi(z)$ versus $z$   in the matter-dominated era for $A= 10^5$  $n>1.5 $ and $k=0$.} 
\label{Figure 17O}
\end{minipage}
\quad
\begin{minipage}[b]{0.5\linewidth}
\includegraphics[width=0.9\textwidth]{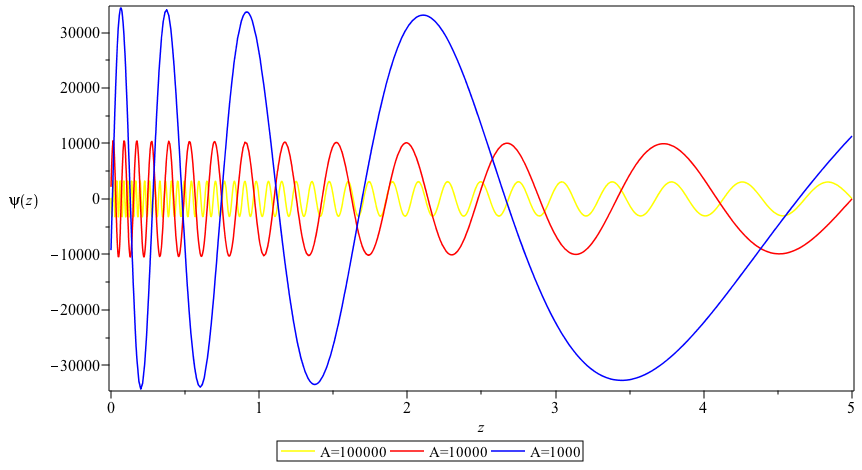}
\caption{$\psi(z)$ versus $z$ in the matter-dominated era for   $n=1.3 $ and $k=0$.} 
\label{Figure 172O}
\end{minipage}
\end{figure}

 For  the case when $k=0$,   Eq. (\ref{psi0003}) is written as
\begin{equation}
\psi''+\dfrac{2}{(1+z)}\psi' - \frac{A}{ (1+z)^{4}}  U(z) \psi=0\;,
\end{equation} 
the general solution is combination for Hypergeometric functions 
\begin{eqnarray}
&&\psi(z)= C_{1} \mathcal{H} \Big([], [\dfrac{(3n-3)}{(4n-3)}], \dfrac{An^{2} B (1+z)^{\frac{-4n+3}{n}}}{(4n-3)^2}\Big)\nonumber\\&&+ C_{2} \dfrac{1}{(1+z)}\mathcal{H} \Big([], [\dfrac{(5n-3)}{(4n-3)}], \dfrac{An^{2}B (1+z)^{\frac{-4n+3}{n}}}{(4n-3)^2}\Big)\;,
\end{eqnarray}
where $B= \dfrac{-2n^{2} }{9} \Big(\dfrac{(-16n^2+26n-6)}{2n^2}+\dfrac{(n-1)(4n-3)}{n^2} +\dfrac{6(n-1)}{n}\Big)\;.$\\
For  $n=1$, we get the general solution  as a combination of Bessel functions as
\begin{equation}
\psi(z)= \dfrac{C_{1}J_{\nu}(a) (1, \dfrac{4\sqrt{A}}{3\sqrt{1+z}})}{\sqrt{(1+z)}}+ \dfrac{C_{2}Y_{\nu}(a) (1, \dfrac{4\sqrt{A}}{3\sqrt{1+z}})}{\sqrt{(1+z)}}\;.
 \end{equation}
This solution corresponds to the solution found in \cite{capozziello2000oscillating}, for the dust case where the cosmological constant  was considered to be $\Lambda=0$. We set the initial conditions, $\psi(z_{in})= 10^{-3}$ and $\psi '(z_{in})=10^{-3}$, where $z_{in}=1100$. The results are presented  in Figs. \ref{Figure 14O} - \ref{Figure 17O}  for a specific choice of the mass $A= 10^5$  and for different values of $n$. Fig. \ref{Figure 14O}, presents the GR results for $n=1$. The probabilities of finding a galaxy of mass $A=10^5$ are slightly increasing with decreasing  the redshift (the amplitudes or the probabilities are approximately constant for most of the cycles) for values of $1\leq n \leq 1.5$ see  Fig. \ref{Figure 16O}, while the probabilities  are decreasing with decreasing  the redshift for $n\geq 1.6$ as shown in Fig. \ref{Figure 17O}.   % the amplitudes or the probabilities are approximately constant for most of the cycles . For a fixed  value of $n$ and different  values of the mass $A$,  see Fig.  \ref{Figure 17O}, we noticed that  the probabilities of finding the galaxies of different masses are increasing with a very small rate with increasing the mass $A$ and the rate of oscillations is increasing with increasing $A$.  We can also noticed  a breaking of homogeneity and isotropy at small scales  with oscillating correlations  between galaxies.
\newpage
\subsection*{\bf{ For  the case when $k=1$ }}
For the case when $k=1$, Eq. (\ref{psi0003}) is written as
\begin{equation}
\psi''+\dfrac{2}{(1+z)}\psi' + \frac{A}{ (1+z)^{4}}(-1-  U(z)) \psi=0\;,
\end{equation} 
and  for $n=1$, the general solution is a combination of Whittaker functions
\begin{equation}
\psi(z)=C_{3}  \mathcal{Y}_{1} \Big( \frac{  2\sqrt{A}}{9}, \frac{1}{2}, \frac{2\sqrt{A}}{(1+z)}\Big)+ C_{4} \mathcal{Y}_{2}  \Big( \frac{  2\sqrt{A}}{9},\frac{1}{2}, \frac{2\sqrt{A}}{(1+z)}\Big)\;.
\end{equation}
We present the numerical solutions  in the following Figs. \ref{Figure 14} - \ref{Figure 144A}. We noticed the oscillatory behaviour  for $n=1$ and $1< n < 2 $ only when the range of redshift begins at $ z \geq 2$ as shown in Figs. \ref{Figure 14} -  \ref{Figure 144a}. While for values of $n \geq1.4 $, oscillations  observed only when the range of redshift begins at $0< z \geq 2$ as shown in Fig. \ref{Figure 144}. We also presented the numerical results for different masses $A$ in Fig. \ref{Figure 144A}
\begin{figure}[h!]
\begin{minipage}[b]{0.45\linewidth}
\includegraphics[width=0.9\textwidth]{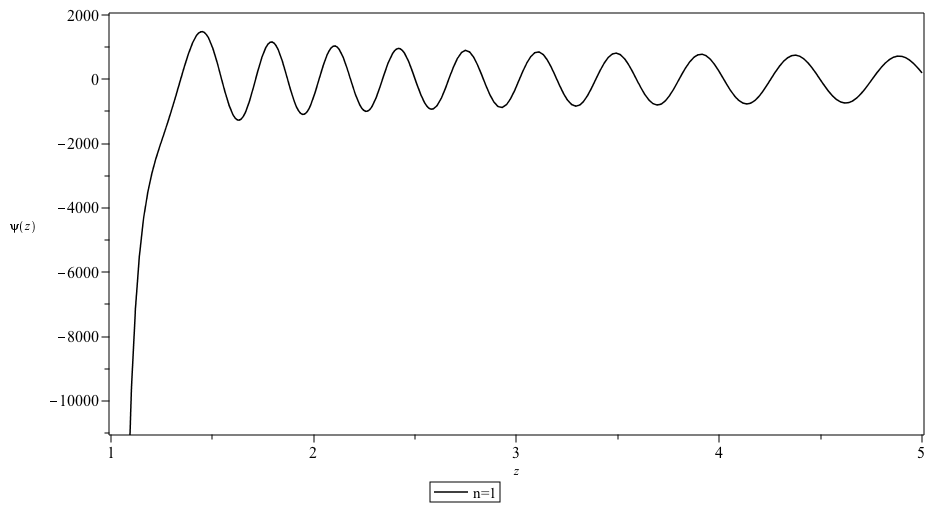}
\caption{$\psi(z)$ versus $z$ in the matter-dominated epoch for $A= 10^5$, $k=1$  and $n=1$.}
\label{Figure 14}
\end{minipage}
\qquad
\begin{minipage}[b]{0.45\linewidth}
\includegraphics[width=1\textwidth]{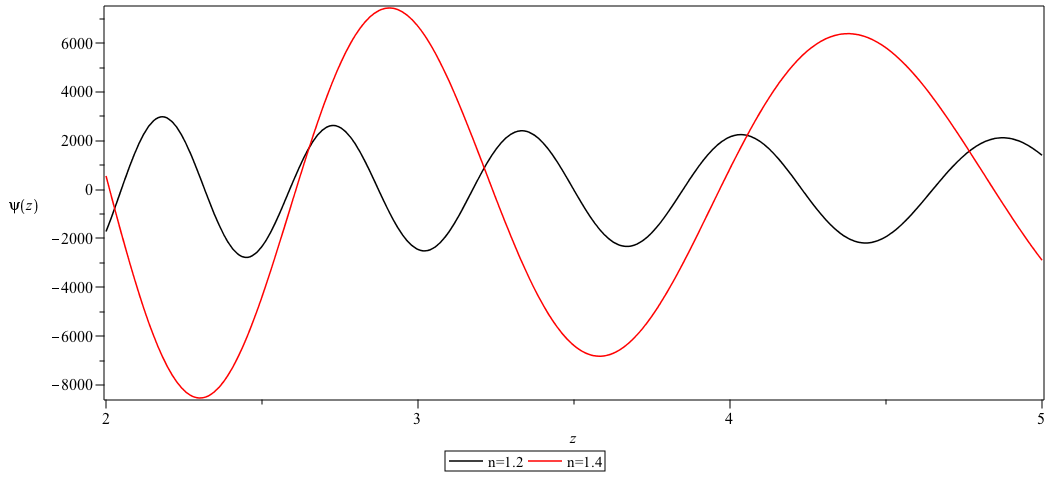}
\caption{$\psi(z)$ versus $z$ in the matter-dominated epoch  for $A= 10^5$,  $k=1$ and $ 1\leq n < 2$.}
\label{Figure 144a}
\end{minipage}
\end{figure}
\begin{figure}[h!]
\begin{minipage}[b]{0.45\linewidth}
\includegraphics[width=0.93\textwidth]{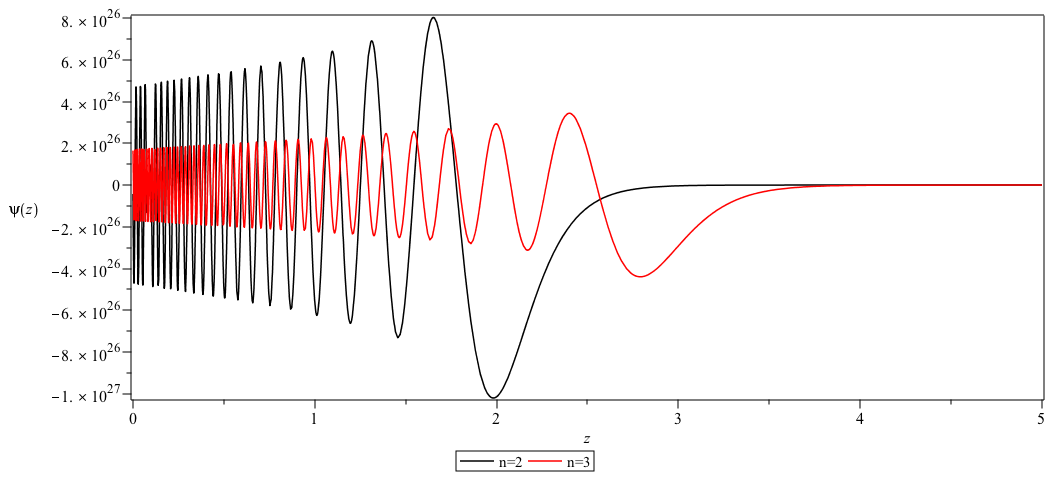}
\caption{$\psi(z)$ versus $z$ in the matter-dominated epoch  for $A= 10^5$,  $k=1$ and $n \geq 2$.}
\label{Figure 144}
\end{minipage}
\qquad
\begin{minipage}[b]{0.45\linewidth}
\includegraphics[width=1\textwidth]{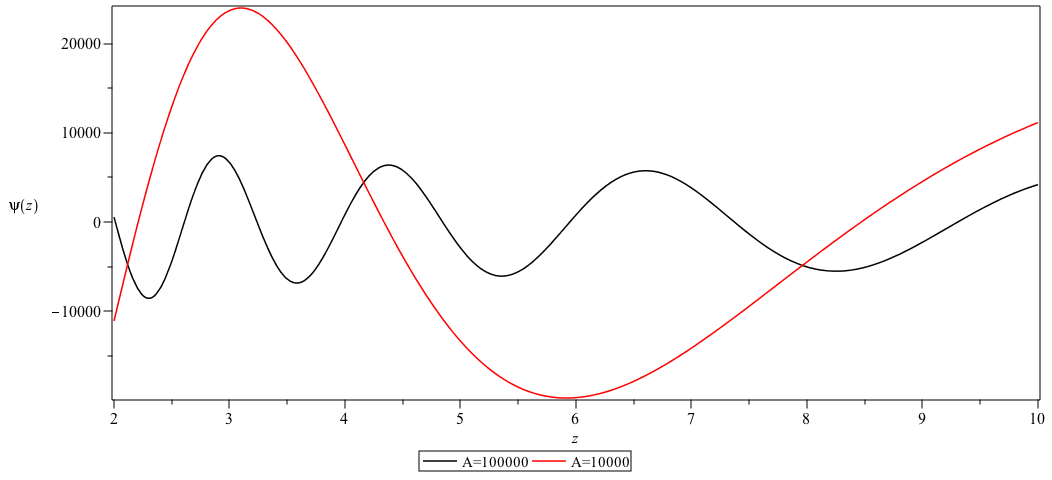}
\caption{$\psi(z)$ versus $z$ in the matter-dominated epoch  for  $k=1$ and $n=1.3$.}
\label{Figure 144A}
\end{minipage}
\end{figure}
\newpage
\subsection*{ \bf{For  the case when $k=-1$ }}
For the case when $k=-1$, Eq. (\ref{psi0003}) is written as
\begin{equation}
\psi''+\dfrac{2}{(1+z)}\psi' + \frac{A}{ (1+z)^{4}}(1-  U(z)) \psi=0\;,
\end{equation} 
and  for $n=1$, the general solution is a combination of Whittaker functions
\begin{equation}
\psi(z)=C_{5}  \mathcal{Y}_{1} \Big(-\frac{2I}{9} \sqrt{A}, \frac{1}{2}, \frac{2I\sqrt{A}}{(1+z)}\Big)+ C_{6} \mathcal{Y}_{2}  \Big(-\frac{2I}{9} \sqrt{A},\frac{1}{2}, \frac{2I\sqrt{A}}{(1+z)}\Big)\;.
\end{equation}
We present the numerical solutions  in the following Figs. \ref{Figure 15} - \ref{Figure 155}.
\begin{figure}[h!]
\begin{minipage}[b]{0.5\linewidth}
\includegraphics[width=0.9\textwidth]{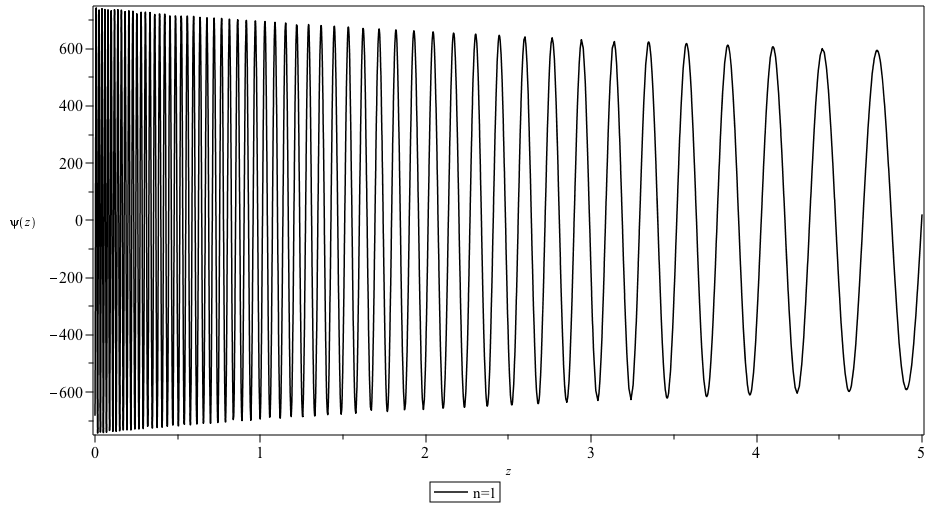}
\caption{$\psi(z)$ versus $z$  in the matter-dominated epoch for $A= 10^5$,  $k=-1$ and $n=1$.} 
\label{Figure 15}
\end{minipage}
\qquad
\begin{minipage}[b]{0.5\linewidth}
\includegraphics[width=0.9\textwidth]{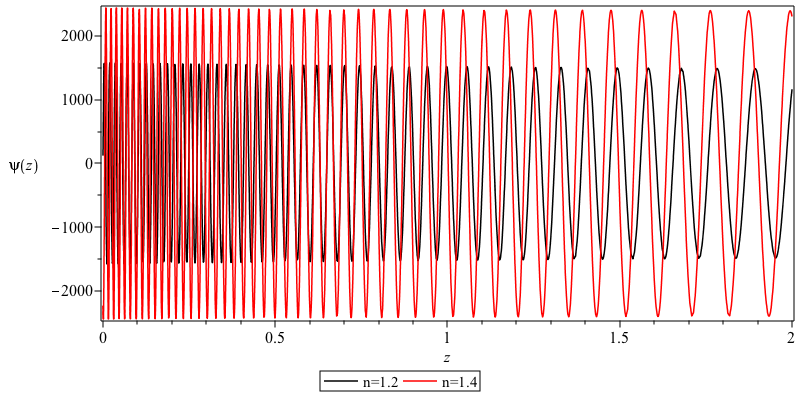}
\caption{$\psi(z)$ versus $z$  in the matter-dominated epoch for $A= 10^5$,  $k=-1$ and $ 1 \leq n \leq 2$.} 
\label{Figure 155}
\end{minipage}
\end{figure}

\section{Conclusions}\label{sec3}
In this paper, we studied the nature of periodic clustering of large-scale cosmic structures in the context of the $f(R)$ gravity theory. Using the $f(R)$-modified Friedmann equations, we were able to construct the cosmological Schr\"{o}dinger-like equation. Following similar works in \cite{capozziello2000oscillating} (and references therein), we considered the particle dynamics of galaxies and calculated the probability of finding them at a certain redshift as eigensolutions of Schr\"{o}dinger equation. We solved the  Schr\"{o}dinger equation for different ranges of $n$ in the  $R^{n}$ model for different equation of state parameters (radiation and pure dust). For the radiation- and dust-dominated epochs in the  flat universe ($k=0$) case, we  got  a combination of hypergeometric functions  as a general solution of the  Schr\"{o}dinger equation. Some of the specific highlights of this work are as follows: For the radiation- and dust-dominated epochs  in the flat universe ($k=0$) and $n=1$, we got a combination of Bessel functions as a general solution (which matches the one for  the GR results  found in \cite{capozziello2000oscillating} for $\psi(a)$) and as we have shown in Figs. \ref{Figure 2}, \ref{Figure 3}, \ref{Figure 16O} and \ref{Figure 17O}, oscillatory behaviour observed for all values of $n$. However, in the radiation-dominated era, the amplitudes of such oscillations decrease with increasing the values of $n$ as in Figs. \ref{Figure 2}, \ref{Figure 3}. For the matter-dominated era, the amplitudes of such oscillations are almost constant for most of the cycles as in Fig. \ref{Figure 16O}.  Numerical solutions have been obtained  for $k=\pm1$ cases for both the radiation- and dust-dominated epochs.  For instance, for $k=-1$, the cosmological solutions are oscillating functions but they are not periodic oscillations (only for a particular range of the redshift $z$) as in Figs. \ref{Figure 5} - \ref{Figure 7} and Figs. \ref{Figure 14} - \ref{Figure 144}. \\
In conclusion, as we have shown in Figs. \ref{Figure 1} to \ref{Figure 155}, for appropriate choices of the initial conditions (and for normalized parameters of the model),  a breaking of homogeneity and isotropy indeed occurs on small scales as depicted by the oscillating correlations between galaxies.
\section*{Acknowledgments}
HS gratefully acknowledges the financial support from the Mwalimu Nyerere African Union scholarship and the National Research Foundation (NRF) free-standing scholarship with a grant number  112544.  AA acknowledges that this work is based
on the research supported in part by the NRF of South Africa with grant number 112131.

\section*{References}

\bibliography{references}
%\renewcommand{\bibname}{References}
%\nocite{*}%\bibliographystyle{revcompchem}
%\bibliographystyle{naturemag}
%\bibliographystyle{amsalpha} %% acm, naturemag, revcompchem
%\bibliographystyle{alpha}
%\bibliographystyle{unsrt}
%\bibliographystyle{apalike}
%\bibliographystyle{amsplain}
%\bibliographystyle{plain}
%\bibliographystyle{ieeetr}
%\bibliographystyle{h-physrev3.bst}
%\bibliographystyle{amsplain}
%\bibliographystyle{abbrv}
%\bibliographystyle{apsrev}
\bibliographystyle{iopart-num}

\end{document}